\documentclass[superscriptaddress,reprint,preprintnumbers]{revtex4-2}  % if error occurs, use TeXLive version 2020

\usepackage{mathtools, amssymb, amsmath, glossaries, multirow, booktabs, comment, graphicx, aeguill}
\usepackage[normalem]{ulem}

\glsunsetall
\newcommand{\zs}{z_{\rm s}}
\newcommand{\zd}{z_{\rm d}}
\newcommand{\rein}{\theta_{\rm Ein}}
\newcommand{\dd}{\mathrm{d}}
\newcommand{\kms}{km s$^{-1}$}
\newcommand{\ms}{M} % stellar mass

\revised{\today}

\begin{document}
\preprint{IPMU22-0067, RIKEN-iTHEMS-Report-22}

\title{Strong Lensing of High-Energy Neutrinos}

\email{yctaak@astro.ucla.edu}

\author{Yoon Chan Taak}
\affiliation{Physics and Astronomy Department, University of California, Los Angeles, CA 90095-1547, USA}

\author{Tommaso Treu}
\affiliation{Physics and Astronomy Department, University of California, Los Angeles, CA 90095-1547, USA}

\author{Yoshiyuki Inoue}
\affiliation{Department of Earth and Space Science, Graduate School of Science, Osaka University, 1-1 Machikaneyama, Toyonaka, Osaka 560-0043, Japan}
\affiliation{Interdisciplinary Theoretical \& Mathematical Science Program (iTHEMS), RIKEN, 2-1 Hirosawa, Saitama 351-0198, Japan}
\affiliation{Kavli Institute for the Physics and Mathematics of the Universe (WPI), The University of Tokyo Institutes for Advanced Study, The University of Tokyo, 5-1-5 Kashiwanoha, Kashiwa, Chiba 277-8583, Japan}

\author{Alexander Kusenko}
\affiliation{Physics and Astronomy Department, University of California, Los Angeles, CA 90095-1547, USA}
\affiliation{Kavli Institute for the Physics and Mathematics of the Universe (WPI), The University of Tokyo Institutes for Advanced Study, The University of Tokyo, 5-1-5 Kashiwanoha, Kashiwa, Chiba 277-8583, Japan}
\affiliation{Theoretical Physics Department, CERN, 1211 Geneva 23, Switzerland}

\begin{abstract}
We consider the effects of strong gravitational lensing by galaxy-scale deflectors on the observations of high-energy (E\,$\gg$\,GeV) neutrinos (HEN). For HEN at cosmological distances, the optical depth for multiple imaging is $\sim 10^{-3}$, implying that while we do not expect any multiply imaged HEN with present samples, next-generation experiments should be able to detect the first such event. We then present the distribution of expected time delays to aid in the identification of such events, in combination with directional and energy information. In order to assist in the evaluation of HEN production mechanisms, we illustrate how lensing affects the observed number counts for a variety of intrinsic luminosity functions of the source population. Finally, we see that the lensing effects on the cosmic neutrino background flux calculation would be negligible by taking kpc-scale jets as an example.

\end{abstract}

\maketitle

\section{Introduction} \label{sec:intro}

The IceCube Neutrino Observatory \citep{Abbasi+09} has been successful in detecting extraterrestrial high-energy (E\,$\gtrsim$\,TeV) neutrinos (HEN) over the past decade \citep{IceCube13}. 
In general, the all-sky distribution of HEN is consistent with isotropy \citep{Aartsen+19}. There have been efforts to pinpoint the sources of these neutrinos; a potential population is blazars, which are expected to create HEN during gamma-ray flares \citep{YoshidaK+22}. 
An outstanding example of this is TXS 0506+056, a blazar that was linked to the neutrino event IceCube-170922A during its flaring period \citep{IceCube18}. 
However, efforts to associate the two individual events are still ongoing \citep{Padovani+18,Banik+19,HwangS+21}. 
In addition, there is some controversy to whether the neutrinos are correlated with potential sources in general \citep{Aartsen+20,Abbasi+22}. 
It is also possible that some of the neutrinos observed by IceCube are produced by cosmic rays accelerated in jets and interacting with photon backgrounds along the line of sight \citep{Essey+10,Essey+11,Kalashev+13,Kochocki+21}. 
Thus, the origin of extraterrestrial neutrinos remains unclear.  

One reason for this uncertainty is the lack of multiplet event detections. The angular resolution of state-of-the-art neutrino telescopes is too large ($\sim 1$ deg), which encumbers the determination of the neutrino sources. Multiple detections from the same source will allow us to better constrain the source position. Unfortunately, with the sensitivity of current or future neutrino telescopes, multiplets are not expected to be frequently detected, especially for sources at high redshifts ($z>2$) \citep{YoshidaS+22}.

Gravitational lensing occurs when spacetime becomes warped around a massive object, and light traveling through this spacetime follows suit. When the source and the deflector are aligned well enough, the resulting images are significantly distorted and multiple images may manifest; this phenomenon is called strong gravitational lensing. Neutrinos are capable of being lensed if the particle speed is relativistic, which allows them to be regarded as photons. 

Strong lensing of distant HEN is a possible explanation for the issues discussed above. Assuming that HEN originate from sources at cosmological distances, if the lensed HEN population dominates over HEN from nearby sources, this can solve both the isotropy enigma and the non-correlation with other object types, such as active galactic nuclei, simultaneously. On sub-arcminute scales, lensed neutrinos will increase the number of detections for certain regions, degrading isotropy. However, the opposite is true for larger scales: distant sources will more or less be isotropic compared to nearby sources, so if the distant neutrinos are boosted by strong lensing, the fraction of neutrinos originating from sources of an isotropic distribution will increase, which in turn enhances isotropy. In addition, lensed HEN allows the detection of multiple neutrinos from the same source, which will assist in constraining the source position and determine the source population of HEN. Unfortunately this is unlikely to be the sole explanation, due to the low strong lensing probability, as we will show below.

There have been previous studies discussing strong lensing of neutrinos. 
\cite{Escribano+01} discuss the possibility of observing lensed neutrinos that pass through the deflector due to their lack of interactions with matter, but since the central image is usually demagnified, the odds of observing such a phenomenon are unrealistic. 
The prospect of using interferometry for lensed neutrinos due to the different paths is demonstrated in \cite{Crocker+04}, but this is also strictly theoretical at this time. 
\cite{Mena+07} consider neutrinos emitted by supernovae that are lensed by objects within the Milky Way, but due to their low masses and consequently small Einstein radii, expectations are tenuous at best. 
\cite{LuoX09} examine strong lensing of neutrinos in general, while focusing on the deviation from geodesics due to the non-zero neutrino mass. 
\cite{Adrian-Martinez+14} examine the effect of magnifications to better constrain the luminosities of several lensed flat-spectrum radio quasars (FSRQs).

The aim of this paper is to investigate how strong lensing by galaxy deflectors affects observed HEN, and discuss the prospect of detecting lensed HEN. We describe the methodology for calculating lensing effects in Section \ref{sec:lensed}, and demonstrate the effects on the observed source luminosity functions in Section \ref{sec:obslf}. 
%In Section \ref{subsec:kpc}, we inspect whether the lensed HEN hypothesis is feasible by studying distant kpc-scale jets as HEN sources as a case scenario. 
We present our conclusions in Section \ref{sec:conc}. The standard $\Lambda$CDM cosmological model with $H_0 = 70$ km s$^{-1}$ Mpc$^{-1}$, $\Omega_{\rm M} = 0.3$, and $\Omega_\Lambda = 0.7$ is used throughout this paper.

\section{Lensed high-energy neutrinos} \label{sec:lensed}

For the lensing analyses in this paper, we can treat HEN as photons. The upper limit for the neutrino mass is about 0.1~eV, so neutrinos with 1~GeV energy have velocities of $\beta=\sqrt{1-\gamma^{-2}}>1-10^{-20}$, and their deviation from paths taken by photons is $\Delta \theta/\theta_{\rm Ein}=1/2\gamma^2<10^{-20}$ \citep{QianL12} (note that $\Delta\theta$ is defined differently from the reference), so the angle difference is insignificant. The presumed energy of 1~GeV is much smaller than the energy regime that is probed in this work, so the path taken by HEN do not deviate from geodesics to the order of less than $10^{-20}$, and thus our approximation of HEN taking photon-like paths is justified.

\subsection{Deflector population} \label{subsec:deflector}

Galaxies are the predominant deflector population for extragalactic sources. We adopt the formulation for the redshift-dependent galaxy velocity dispersion function (VDF) introduced by \cite{Mason+15}, which is summarized here. The local galaxy VDF measured by observing SDSS galaxies \citep{ChoiY+07} takes the form of a modified Schechter function as follows: 
\begin{equation}
\begin{aligned}
\Phi_\sigma(\sigma,z=0) \:\dd\sigma = &\Phi^*_\sigma \:\Big(\dfrac{\sigma}{\sigma^*}\Big)^{\alpha_\sigma} \:\exp\Big[-\Big(\dfrac{\sigma}{\sigma^*}\Big)^{\beta_\sigma}\Big]\\
&\times \:\dfrac{\beta_\sigma}{\Gamma(\alpha_\sigma/\beta_\sigma)}\: \dfrac{\dd\sigma}{\sigma}\\
\end{aligned}
\end{equation}
where $\sigma$ is the velocity dispersion, $\Phi^*_\sigma=8.0\times10^{-3} \:h^3 \: {\rm Mpc}^{-3}$ with $h\equiv H_0/(100$ \kms $\:{\rm Mpc}^{-1}) =0.7$, $\sigma^* = 161$ \kms, $\alpha_\sigma=2.32$, $\beta_\sigma=2.67$, and $\Gamma$ is the gamma function. Previous studies \citep{Wyithe+11} have used this constant galaxy VDF with respect to redshift because it does not appear to evolve up to $z\sim1$, where a significant portion of the deflector population lies. However, we intend to investigate the lensing effects for sources potentially at high redshift, so such an evolution is required for an accurate analysis. 

We start with the stellar mass function from \cite{Muzzin+13}, which is a Schechter function in the following form: 
\begin{equation}
\Phi_M(\ms) \:\dd \ms = \Phi^*_M \:\Big(\dfrac{\ms}{\ms^*}\Big)^{1+\alpha_M} \:\exp\Big[-\Big(\dfrac{\ms}{\ms^*}\Big)\Big] \:\dfrac{\dd\ms}{\ms},\\
\end{equation}
where $\ms$ is the stellar mass, $\Phi^*_M$ is the normalization, $\ms^*=10^{11.06}M_\odot$ is the characteristic stellar mass, and $\alpha_M=-0.54$ is the low-mass-end slope. As is shown by \cite{Mason+15},  the evolution of the three parameters can be linearly parameterized with sufficient approximation. Nonetheless, the evolution for $\ms^*$ is ignored since its effect on the resulting LF is negligible, and likewise for $\alpha_M$ because the low-mass end is mostly irrelevant to lensing arguments, so only a linear evolution of $\Phi^*_M(z)=3.75 \times 10^{-3} \times (1+z)^{-2.46}$ Mpc$^{-3}$ is applied. 

It is well known that $\ms$ and $\sigma$ follow a linear correlation in the form of $\log (\sigma/$\kms$) = p[\log (\ms/M_\odot)-11]+q$ with $p=0.24$ and $q=2.32$ \citep[e.g.,][]{Auger+10}, so $\ms \propto \sigma^{1/p}$. However, at higher redshifts, massive galaxies have larger velocity dispersions when compared to their local counterparts at fixed stellar mass , and we can model the evolution as $\sigma=\sigma_0\:(1+z)^{k_\sigma}$, where $\sigma_0$ is the velocity dispersion expected from the $\ms-\sigma$ correlation at $z\sim0$, and $k_\sigma=0.20$ is the strength of the evolution from \cite{Mason+15}. Therefore we can expect an evolution of the $\ms-\sigma$ correlation as $\ms\propto[\sigma/(1+z)^{k_\sigma}]^{1/p}$, and the evolving VDF becomes 
\begin{equation}
\begin{aligned}
\Phi_\sigma(\sigma,z) \:\dd \sigma = &\:\Phi^*_M(z) \:\Big(\dfrac{\sigma}{\sigma^*_0\:(1+z)^{k_\sigma}}\Big)^{(1+\alpha_M)/p}\\
&\times \:\exp\Big[-\Big(\dfrac{\sigma}{\sigma^*_0\:(1+z)^{k_\sigma}}\Big)^{1/p}\Big] \:\dfrac{1}{p} \:\dfrac{\dd\sigma}{\sigma},\\
\end{aligned}
\end{equation}
where $\sigma^*_0=216$ \kms{} is the conversion from $\ms^*$ using the local $\ms-\sigma$ correlation.

\begin{figure*}
%\makebox[\textwidth][l]{\includegraphics[width=0.5\textwidth]{3-3_vdf.eps}}
\makebox[\textwidth][l]{\includegraphics[width=\textwidth]{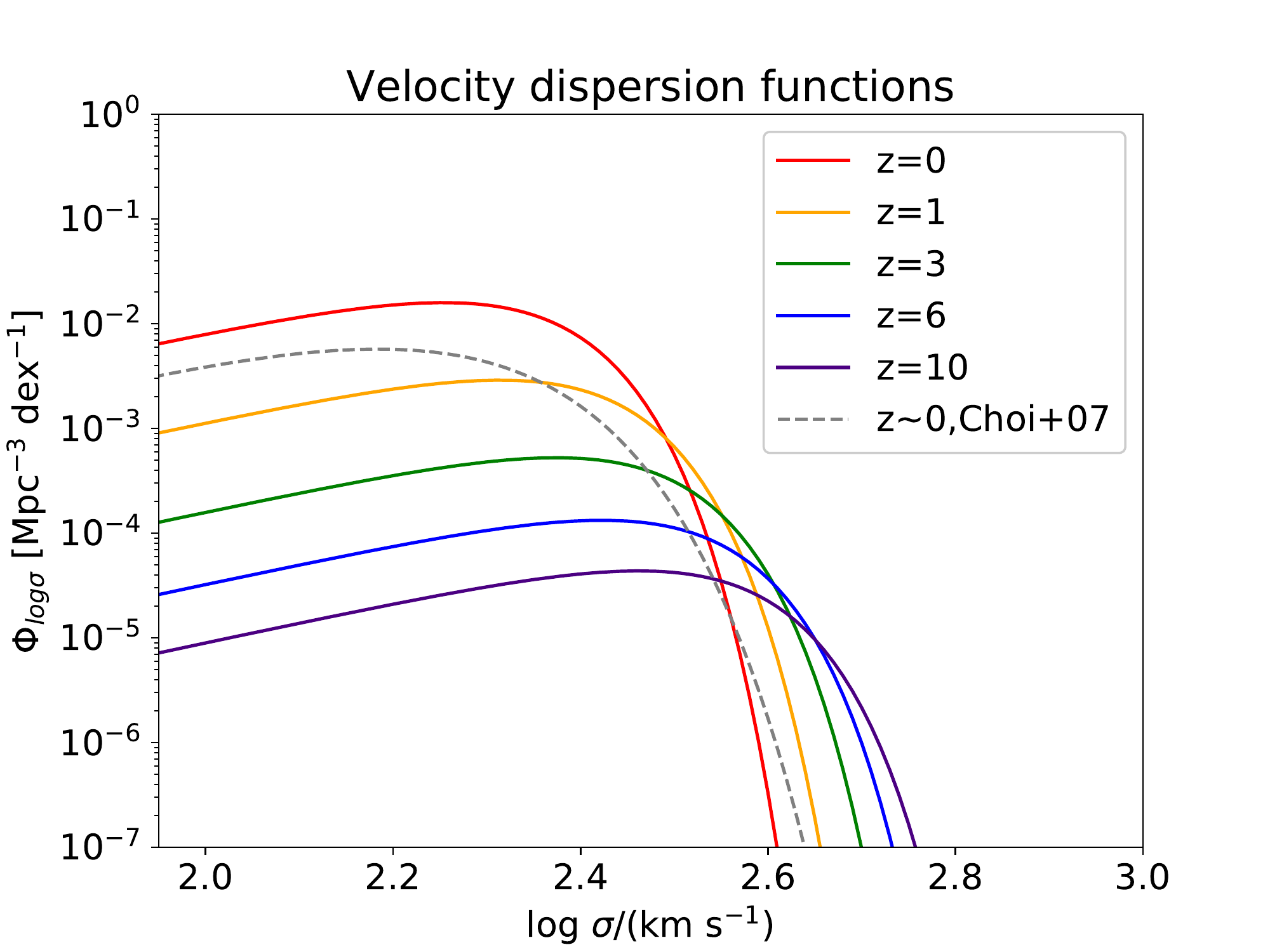}}
\caption{VDFs for various redshifts. Red, yellow, green, blue and violet solid lines correspond to the VDFs at $z=0,1,3,6,$ and 10, respectively, following the evolution discussed in Section \ref{subsec:deflector}. The gray dashed line is for the local VDF from SDSS \citep{ChoiY+07}.
}
\label{fig:vdf}
\end{figure*}

The evolving VDF is shown in Figure~\ref{fig:vdf}, along with the local VDF. The VDF continuously shifts towards larger velocity dispersions following the evolution of the $\ms-\sigma$ correlation, but simultaneously the normalization decreases due to the evolution of $\Phi_M^*$. The local VDF lies somewhere between the VDFs for $z=0$ and 1, as expected. 

\subsection{Optical depth} \label{subsec:optdepth}

The optical depth ($\tau$) can be roughly interpreted as the probability of a source at redshift $\zs$ to be multiply imaged, which is equivalent to the probability of a light ray passing through the lensing cross-section of any deflector. In equation form, this is 
\begin{equation}
\tau (\zs) = \int^{\zs}_{0} d\zd \:\dfrac{dV}{d\Omega \:d\zd} \:\dfrac{dN}{dV} \:\Omega\\
\end{equation}
where $\zd$ is the deflector redshift, $V$ is the comoving volume, $N$ is the number of deflectors, and $\Omega=\pi\rein^2$ is the solid angle corresponding to the cross-section for strong lensing by a singular isothermal spherical mass distribution, with $\rein$ being the Einstein radius. From \cite{Hogg99}, the comoving volume is expressed as 
\begin{equation}
d V(z) = \dfrac{c}{H_0} \:\dfrac{(1+z)^2 \:d_{\rm A}^2}{E(z)} \:d\Omega \:d z\\
\end{equation}
where $d_{\rm A}$ is the angular diameter distance, and $E(z)=[\Omega_M(1+z)^3+\Omega_\Lambda]^{1/2}$ is the dimensionless Hubble parameter. Rearranging for the deflector VDF $\Phi_\sigma = d N/d V d\sigma$, the optical depth becomes 
\begin{equation}
\begin{aligned}
\tau (\zs) = \int^{\zs}_0 d\zd \int d\sigma &\:\Phi_\sigma(\sigma,\zd) \:\dfrac{c}{H_0} \:\dfrac{(1+\zd)^2 \:d_{\rm od}^2}{E(\zd)} \\
& \times \:\pi \rein^2(\sigma,\zd,\zs)
\end{aligned}
\end{equation}
where $\Phi(\sigma,\zd)$ is the galaxy VDF at $\zd$, $d_{\rm od}$ is the angular diameter distance from the observer to the deflector, and $\rein(\sigma,\zd,\zs)$ is the Einstein radius for a deflector with velocity dispersion $\sigma$ at $\zd$ and source at $\zs$. 

\begin{figure*}
%\makebox[\textwidth][l]{\includegraphics[width=0.5\textwidth]{3-4_tau.eps}}
\makebox[\textwidth][l]{\includegraphics[width=1\textwidth]{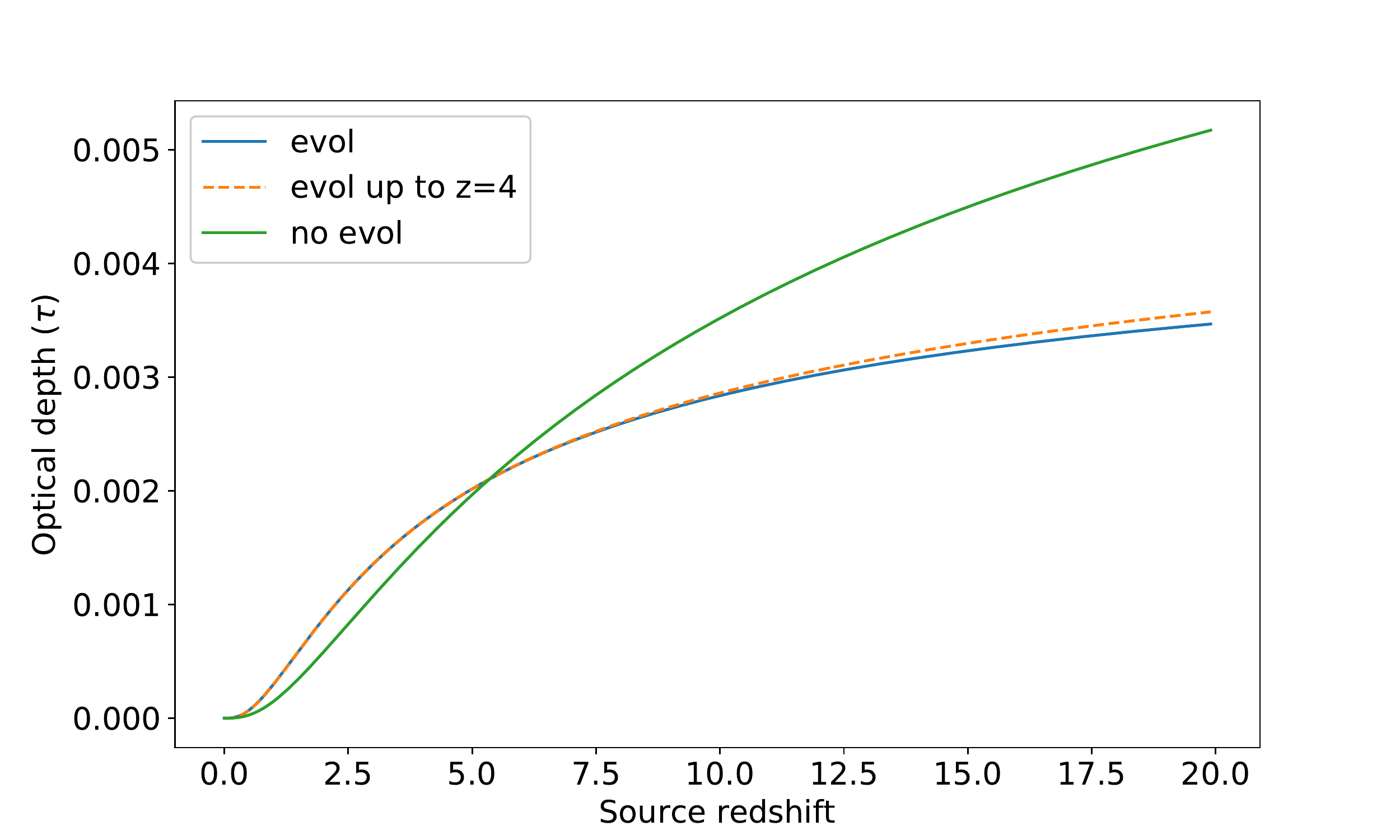}}
\caption{Redshift dependence of optical depths for various scenarios. The green solid line denotes the optical depth for a non-evolving VDF, and the blue solid and orange dashed lines correspond to the optical depths for the VDF evolution described in Section \ref{subsec:deflector} applied to the full redshift range and up to $z=4$, respectively. 
}
\label{fig:tau}
\end{figure*}

Figure \ref{fig:tau} shows the redshift dependence of the optical depth for several evolution scenarios. The first scenario assumes no evolution of the local VDF from \cite{ChoiY+07}. The second one employs the VDF evolution shown in Sec. \ref{subsec:deflector}, and the final model follows the evolution only up to $z=4$, which is the redshift range probed by \cite{Muzzin+13}. We can see that the optical depths are similar for all three scenarios. Delving into details, compared to the no-evolution scenario, the evolution models have larger $\tau$ at lower redshifts and smaller $\tau$ at higher redshifts, with the transition occurring around $z\simeq5$. Also, the latter two scenarios yield almost identical results beyond $z=4$; this is because the number density of galaxies at $z\gtrsim4$ is insignificant to strong lensing compared to those at lower redshifts, and thus verifies that galaxies at lower redshifts are the dominant deflector population for even the most distant sources. We use the second scenario (full evolution) for future discussions.

\subsection{Time delay distributions}

Time delays between multiple lensed images are crucial information for determining whether a source is strong lensed. The time delay of a system depends on the source redshift as $\Delta\tau \propto d_{\rm os}/d_{\rm ds}$, where $d_{\rm os}$ and $d_{\rm ds}$ are the angular diameter distances from the observer/deflector to the source, respectively \citep{Treu+16}. As long as the source is not too close to the deflector, this ratio does not vary significantly. Therefore, we assume that time delays of lensed neutrinos are comparable to those of other sources with the same deflector population, and resort to exploiting time delay distributions from simulations and observations of galaxy-scale lenses in the literature. 

The time delays shown in Figure 8 of \cite{OguriM+10} are for simulated lensed quasars and supernovae. They use the SDSS VDF from \cite{ChoiY+07}, so the deflector population is very similar to the one used in this work. For two-image systems, time delays range from day to year scales; for quads, the delay between the first and last images are of the same scale, although shorter time scales are expected for the images in between. About 70\% of the systems display time delays between 10 and 120 days \citep{LiaoK+15}. Another mock catalog of lensed quasars \citep{YueM+22} also presents a similar distribution. 

\cite{Schmidt+23} have predicted time delays for 30 observed quad quasars using lens modeling. Their Table~8 shows that $19/30=63.3\%$ of the systems are expected to have time delays between 10 and 120 days, and this fraction becomes $11/16=68.8\%$ if we only consider systems with confirmed deflector redshifts. Based on this agreement between the time delay distributions of simulations and observations, we conclude that lensed neutrinos will exhibit time delays of day to year scales, with most of them lying between 10 and 120 days.

\subsection{Future detection of lensed neutrinos} \label{lensednu}

As is seen from Figure \ref{fig:tau}, the optical depth of galaxies for sources at $z\sim1$--3 is $\tau(\zs=2)\sim10^{-3}$, and that for more distant sources increases to the order of 3$\times10^{-3}$, indicating that roughly one in 300--1000 HEN should be lensed. Considering that the number of currently detected HEN is $\sim100$ \citep{Abbasi+21}, it is unlikely that lensed neutrinos with these energies would have been detected with current instrumentation. 

The next-generation detectors, IceCube-Gen2 \citep{Clark+21} and KM3NeT \citep{Adrian-Martinez+16}, are expected to increase the number of neutrinos by an order of magnitude. Thus, it is natural to postulate the detection of at least one lensed neutrino in the near future, and contemplate what such a detection may look like. Since lensing does not alter the neutrino energy and only bends the particle path slightly, two or more detections at the same neutrino energy and almost identical direction in the sky with time delays of days to years will be a strong candidate for a lensed neutrino. However, the expected angular resolution is still at the (sub-)degree level \citep{Aartsen+19}, which is much larger than arcsecond-scale Einstein radii, so it will be impossible to resolve the multiple images. Therefore, confirmation of its lensed nature will have to rely on spatial and energy coincidence and a small delay in arrival time. 

We note that multiplet events originating from a single source may act as contaminants. Sources such as core-collapse SNe and tidal disruption events may emit several neutrinos within time scales comparable to the day-to-year time delays discussed above \citep{YoshidaS+22}. Thus, the rejection of these contaminants will depend on the energy resolution of the experiments; if the difference in energies of multiple neutrinos is larger than the instrumental resolution, then lensing can be excluded. 

When predicting the number of observed neutrinos, it is important to consider the detection efficiency of neutrino detectors, i.e., the fraction of neutrinos entering the detector that are perceived by it. We can estimate this by comparing the number of HEN emitted by sources to those detected by IceCube, using simple order-of-magnitude calculations. The number of HEN emitted by a blazar flare can be approximated by $f_\nu\times A\times \Delta t$, where $f_\nu$ is the neutrino number-flux, $A$ is the effective area of the detector, and $\Delta t$ is the duration of the blazar flare. According to \cite{Oikonomou+19}, based on a sample of bright blazars, the typical blazar flare is observed with IceCube with an effective area of 10$^6$ cm$^2$ and has a duration of 10$^6$ s. So for the blazar flare to emit at least one neutrino, the threshold neutrino number-flux is 10$^{-12}$ neutrinos~cm$^{-2}$~s$^{-1}$, which corresponds to a neutrino flux of 2$\times$10$^{-9}$ erg~cm$^{-2}$~s$^{-1}$ for PeV-scale neutrinos. With an optimistic assumption that the PeV neutrino flux is similar to the GeV gamma-ray flux, the threshold event rate is 10$^{-7}$ events~cm$^{-2}$~s$^{-1}$ at 10~GeV. The number of flares satisfying this criterion is of the order of 100 \citep{Abdollahi+17}, which is equivalent to the number of neutrinos detected by IceCube; this implies that the detection efficiency is approximately 1.

It is possible to expand this argument to IceCube-Gen2, which is expected to have an increase of the effective area by a factor of 10 \citep{Aartsen+21}, so the threshold event rate for 10~GeV should be lowered by the same factor to 10$^{-8}$ events~cm$^{-2}$~s$^{-1}$. Nearly all of the 1994 flares in \cite{Abdollahi+17} are above this flux, so our previous assumption of the number of neutrinos to be increased by a factor of 10 is valid, and we expect several lensed neutrinos to be found with detectors in the next generation.

A potential candidate of lensed neutrinos worthy of note is PKS 1830-211, which is a lensed FSRQ \citep{Lovell+96,Lovell+98b,Lovell+98a} with one of the highest neutrino fluxes across the sky \citep{YoshidaK+22}. HEN from this source will be detectable with KM3Net or IceCube and its upgrade.

\section{Observed source luminosity functions} \label{sec:obslf}

In this section, we discuss the effects of strong lensing on the observed source luminosity functions. 

\subsection{Magnification bias and its effects on the observed LF} \label{subsec:magbias}
As discussed in Section \ref{subsec:optdepth}, the optical depth is equivalent to the probability of a source to be strongly lensed. However, lensing affects number counts (e.g., LFs) in a more complicated mechanism; the luminosity and solid angle of the source are both boosted also. These two effects indicate that for a single value of the magnification $\mu$, the purely lensed portion of a source LF can be expressed as 
\begin{equation}
\Phi_{\rm lensed} \propto \dfrac{N(L)}{\Omega} = \dfrac{\tau N_0(L/\mu)}{\mu\Omega_0} \propto \dfrac{\tau}{\mu}\:\Phi_0\Big(\dfrac{L}{\mu}\Big)
\end{equation}
where $L$ is the source luminosity and $\Phi_0$ is the intrinsic source LF. Thus, the observed total LF is the sum of the unlensed and purely lensed portions of the source LF, or 
\begin{equation}
\Phi_{\rm obs} = (1-\tau)\:\Phi_0(L) + \dfrac{\tau}{\mu}\:\Phi_0\Big(\dfrac{L}{\mu}\Big).
\end{equation}

In reality, $\mu$ depends on the angular position of the source from the deflector, so the latter term is modified to take the probability distribution of $\mu$, or $p(\mu)$, into account. Finally, a demagnification $\mu_{\rm demag} = (1-\overline{\mu}\tau)/(1-\tau)$ is introduced, with $\overline{\mu}$ being the mean magnification for the multiply-imaged region, so that the mean magnification of the full sky is unity, and the observed LF becomes 
\begin{equation}
\Phi_{\rm obs} = (1-\tau)\:\dfrac{1}{\mu_{\rm demag}}\:\Phi_0\Big(\dfrac{L}{\mu_{\rm demag}}\Big) + \tau\int\:\dfrac{\dd p}{\mu}\:\Phi_0\Big(\dfrac{L}{\mu}\Big).
\end{equation}
For singular isothermal spherical deflectors, when considering only the brighter image, $dp(\mu)/d\mu = 2/(\mu-1)^3$, so $\overline{\mu} = \int^\infty_2 \mu \:dp = 3$ and $\mu_{\rm demag} = (1-3\tau)/(1-\tau)$.

\subsection{Observed LFs for several models}

\begin{figure*}
\includegraphics[width=1\textwidth]{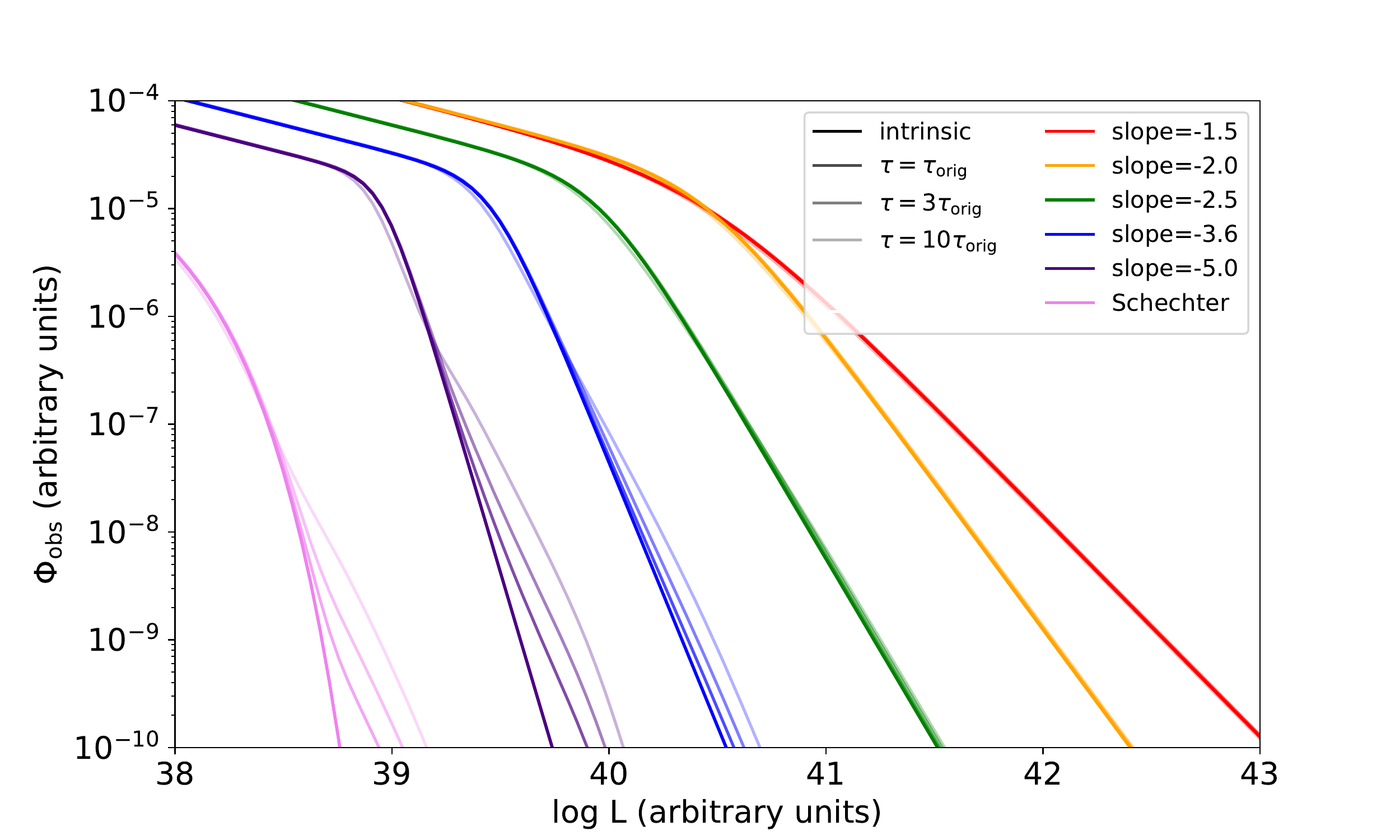}
\caption{Observed LFs at $\zs=20$ for various optical depths and bright-end slopes. LFs with bright-end slopes of $\alpha=-1.5, -2, -2.5, -3.6, -5.0$, and a Schechter function are shown in red, yellow, green, blue, indigo and violet, respectively. The curves are shifted in the x-direction for each of the slopes for clarity, and the choice of $\Phi^*, L^*$ and $\beta$ are arbitrary. The intrinsic LFs are shown with thick lines, and the three observed LFs with different optical depths (with $\tau_{\rm orig}$ indicating the optical depth at $\zs=20$ from Section \ref{subsec:optdepth}) are shown with three corresponding lines for each intrinsic LF with increasing transparency. 
}
\label{fig:obslf}
\end{figure*}

In Figure \ref{fig:obslf}, we plot the observed LFs for $\zs=20$ while varying two parameters; the optical depth and the bright-end slope of the LF. First, we can see that increasing the optical depth generates a larger boost to the bright end of the LF. This is as expected, since more deflectors will enhance the lensing probability, and thus create a more significant effect. Unfortunately, the optical depth at $\zs=20$ for the galaxy deflector population from Section \ref{subsec:deflector} ($\tau_{\rm orig}$) does not augment the LF substantially, and as was seen in Figure \ref{fig:tau}, the optical depth continuously increases with redshift, indicating that for nearer, more realistic sources, the effect is even smaller. In addition, different evolution scenarios do not increase the optical depth by more than factors of several. 

Second, steeper LFs are more susceptible to lensing, and a Schechter function is affected the most. The bright-end slope of the LF is critical in determining whether strong lensing boosts the LF, in that it needs to be steeper than $-2$ for the boost to occur \citep{Blandford+92}. This is demonstrated in Figure \ref{fig:obslf}; the LFs with bright-end slopes of $-1.5$ and $-2$ exhibit no visible change due to lensing, while for steeper slopes the effects of lensing begin to take place. We emphasize that the results shown in this section are applicable not just to neutrinos, but all types of sources of relativistic particles or photons.

\subsection{kpc-scale jets as neutrino sources} \label{subsec:kpc}

In this section, we focus on a specific population for neutrino generation, namely kpc-scale jets, and discuss the possibility of these being neutrino sources. 

As discussed in Section \ref{sec:intro}, the identity of HEN sources is ambiguous. They are likely to be located at large distances to explain the isotropic distribution, so potential populations of HEN sources include quasars and GRBs \citep[see][for reviews]{Ahlers+18,MuraseK+19}. 

Recent observations suggest that high-energy (i.e., TeV) $\gamma$-rays are emitted from the kpc-scale jets of blazars \citep{HESS20}. This discovery indicates the existence of high-energy cosmic rays in the kpc-scale jets. The interaction of particles from relativistic jets and the interstellar medium is a conceivable mechanism of neutrino generation. Thus, we can postulate that $pp$ interactions caused by the collision of jet protons with interstellar gas particles generate neutrinos, and this is a potential source population of distant HEN.    

\subsubsection{Source population} \label{subsubsec:source}
The sources of interest are neutrinos, so we need the intrinsic (i.e., unlensed) neutrino LF ($\nu$LF) as a function of redshift. Unfortunately the number density of neutrino sources is not well understood. Therefore we illustrate our lensing formalism by making a series of assumptions to estimate the $\nu$LF. We stress that our purpose is to elucidate the formalism, not identify the source of HENs.

As an example, we consider protons from AGN jets colliding with ambient gas particles as the major source of neutrinos. So the neutrino luminosity can be obtained from the jet power, which in turn can be deduced from the radio luminosity. 

Thus, we begin with the AGN radio LF at 325 MHz, which is provided by the Galaxy and Mass Assembly (GAMA) survey \citep{Prescott+16} as a double power-law function in the form of 
\begin{equation}
\Phi_{\rm r_1} (L_{\rm r_1}, z=0) \:\dd L_{\rm r_1}= \dfrac{\Phi^*_{\rm r_1}}{(L^*_{\rm r_1}/L_{\rm r_1})^{\alpha_{\rm r}}+(L^*_{\rm r_1}/L_{\rm r_1})^{\beta_{\rm r}}} \:\dd L_{\rm r_1},\\
\end{equation}
where $L_{\rm r_1}$ is the 325 MHz luminosity, $L^*_{\rm r_1}$ is the break luminosity, $\Phi^*_{\rm r_1}$ is the normalization at the break, and $\alpha_{\rm r}$ and $\beta_{\rm r}$ are the bright-end and faint-end slopes, respectively. Two evolutionary scenarios, the pure luminosity and pure density evolutions, were considered; the former postulates that galaxies have undergone a constant decrease in their luminosities without changing their number densities (e.g., mergers), whereas the latter presumes a continuous decrease in their number densities with no change in the break luminosity. For the PLE scenario, the LF evolution is parameterized as 
\begin{equation}
\Phi_{\rm r_1} (L_{\rm r_1},z) = \Phi_{\rm r_1} (L_{\rm r_1}/(1+z)^{k_{\rm r}},z=0),\\
\end{equation}
whereas the parameterization for the PDE case is 
\begin{equation}
\Phi_{\rm r_1} (L_{\rm r_1},z) = \Phi_{\rm r_1} (L_{\rm r_1},z=0) \:(1+z)^{k_{\rm r}},\\
\end{equation}
with $k_{\rm r}$ representing the evolution strength. The redshift-dependent functional forms for both scenarios are shown in Table \ref{tbl:radiolf}. Although this LF evolution is derived only for radio AGNs at $z<0.5$, we extrapolate this to higher redshifts for lack of better data. 

Next we translate the 325 MHz luminosity to the neutrino luminosity ($L_\nu$) using several relations. First, assuming a radio spectral index of $\alpha_f=0.8$, $L_f \propto f^{-\alpha_f}$, so $L_{\rm r_2} = 1.7 \:L_{\rm r_1}$, where $L_{\rm r_2}$ is the radio luminosity at 151 MHz. 

The second relation is an empirical one between the 151 MHz luminosity and time-averaged jet power \citep{Willott+99,InoueY+17} expressed as 
\begin{equation}
P_{\rm jet} = 9.5 \times 10^{46} \Big(\dfrac{f}{10}\Big)^{3/2} \Big(\dfrac{L_{\rm r_2}}{\rm 4\pi \times 10^{28} \:W \:Hz^{-1}}\Big)^{6/7} {\rm erg \:s^{-1}}, \\
\end{equation}
where $f$ is a factor accounting for various errors in the modeling procedure, and assumed to be 10 in this work.

The final equation is the combined result of several assumptions. We assume that roughly 10\% of the jet power consists of protons, and the neutrino efficiency, $f_{pp}$, is calculated as 
\begin{equation}
f_{pp} = \dfrac{t_{\rm dyn}}{t_{pp}} = \dfrac{l_{\rm jet}/c}{1/(n_{\rm gas}\:\sigma_{pp}\:\kappa\:c)} = \kappa \:\sigma_{pp} \:n_{\rm gas} \:l_{\rm jet}\\
\end{equation}
where $\kappa$ is the inelasticity, which is the efficiency of pion production for $pp$ interactions (i.e., the fraction of kinetic energy transferred to the pion), $\sigma_{pp}$ is the cross-section for $pp$ interactions, $n_{\rm gas}$ is the gas density of the interstellar medium, and $l_{\rm jet}$ is the distance from the AGN to the end of the jet, which is the distance traveled by the protons \citep{InoueY+19}. Assuming typical values of $\kappa=0.17$ (for a single neutrino flavor), $\sigma_{pp} \approx 30 \:{\rm mb}$, $n_{\rm gas} \approx 1 \:{\rm cm}^{-3}$, and $l_{\rm jet} \approx 1 \:{\rm kpc}$, $f_{pp} \approx 5 \times 10^{-5}$, so the luminosity for all three flavors is 
\begin{equation}
L_\nu = 0.1 \:f_{pp} \:P_{\rm jet} \approx 5\times10^{41} \:\Big(\dfrac{L_{\rm r_2}}{\rm 4\pi \times 10^{28}\: W \:Hz^{-1}}\Big)^{6/7} \:{\rm erg \:s}^{-1}.\\
\end{equation}

The conclusive translation between $L_{\rm r_1}$ and $L_\nu$ becomes 
\begin{equation}
L_\nu = 5 \times 10^{41} \: \Big(\dfrac{0.13\:L_{\rm r_1}}{\rm 10^{28} \:W \:Hz^{-1}}\Big)^{6/7} {\rm erg \:s^{-1}},\\
\end{equation}
so using this relation, we can convert the 325 MHz LF to the $\nu$LF. Note that due to the power dependence, the slopes of the $\nu$LF should be 7/6-ths of the radio LF slopes, so the bright-end slope of the $\nu$LF is about $-3.6$ in this case.

\subsubsection{Observed $\nu$LF}
\begin{figure*}[pt]
\makebox[\textwidth][l]{\includegraphics[width=1\textwidth]{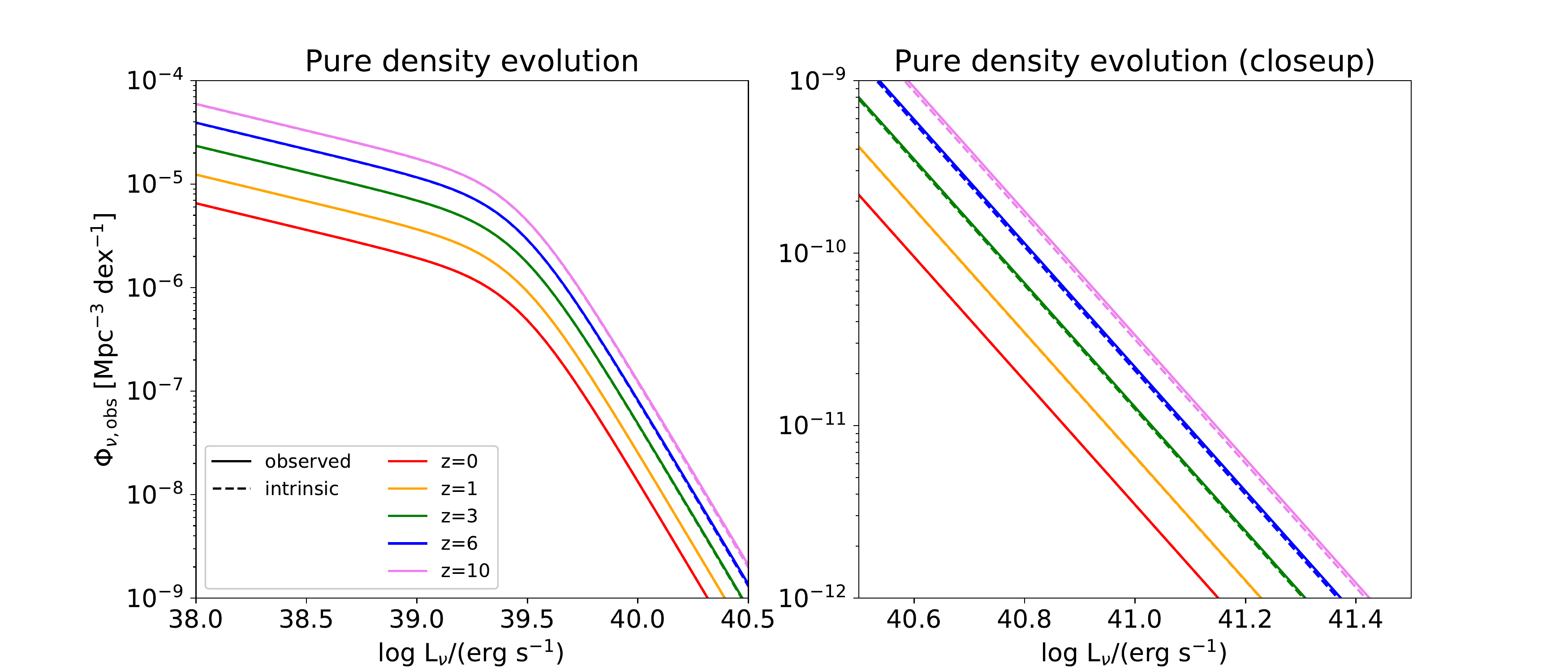}}
\caption{Observed $\nu$LFs for the PDE model described in Section \ref{subsubsec:source}. Dashed lines indicate the intrinsic LFs, while solid lines represent the observed LFs. Red, yellow, green, blue and violet lines are for LFs at $z=0, 1, 3, 6$ and 10, respectively. The right panel is simply a magnified version. 
}
\label{fig:obslf2}
\end{figure*}

Figure \ref{fig:obslf2} shows the observed $\nu$LF for the PDE scenario of the radio LF described in Section \ref{subsubsec:source}. We can see that the effects of strong lensing on the $\nu$LF exists, because the bright-end slope is steeper than $-2$, but the differences between the intrinsic and observed $\nu$LFs are virtually indistinguishable. This result is compatible with what is expected from Section \ref{sec:obslf}; the slope corresponds to the blue lines in Figure \ref{fig:obslf}, but the sources are located at redshifts much less than $\zs=20$, so their optical depths are smaller than $\tau_{\rm orig}$, and effects on the LF are minimal. A steeper slope of the bright-end of the intrinsic LF and/or a more distant source population is required for a prominent boost to the observed LF. The PLE model is expected to show near-identical results, since the bright-end slopes are similar. 

An issue for consideration is that the bright-end slope of the radio LF is quite uncertain; many radio LFs in the literature usually have bright-end slopes between $-1$ and $-2$, which translates to bright-end slopes for the intrinsic $\nu$LF between $-1.17$ and $-2.33$. This is shallower than the bright-end slope used here, so even when using these alternative radio LFs, the results will be minimally affected by strong lensing, not only since the slopes are shallower than what is used in the previous section, but also because they are shallower than or marginally steeper than the threshold of $-2$. 

Thus, we conclude that lensing effects on the observed LF are negligible for the model taken as an example.

\section{Conclusion} \label{sec:conc}

In this paper, we discuss how the detection of extragalactic high energy neutrinos are affected by strong lensing. First, we show that the optical depth of galaxies as a function of redshift is roughly consistent over several evolution scenarios, and that $\tau(\zs\approx2) \sim 10^{-3}$ and $\tau(\zs\gtrsim10) \sim 3\times10^{-3}$. Based on these calculations, we predict that at least one lensed neutrino will be discovered in the near future, and suggest several means of identifying them, such as their expected time delays, along with coincidence in energy and position in the sky. 

In addition, we examine how source LFs are altered due to strong lensing effects, and illustrate visually that bright-end slopes steeper than $-2$ are required for the LFs to be boosted by lensing, and that LFs with steeper slopes are augmented more. Finally, kpc-scale jets are investigated in detail as an example of potential neutrino sources, and we demonstrate that changes to the observed $\nu$LFs are insignificant. 

To conclude, the detection of lensed neutrinos is at hand, and this paper provides some tools and guidance on how to identify and confirm them.

\acknowledgments

We thank Charlotte A. Mason for discussions regarding the evolution of the galaxy VDF. 

Y.C.T. is supported by the Basic Science Research Program through the National Research Foundation of Korea (NRF) funded by the Ministry of Education (grant number 2021R1A6A3A14044070). 
A.K. was supported by the U.S. Department of Energy (DOE) Grant No. DE-SC0009937, by the Simons Foundation, by the World Premier International Research Center Initiative (WPI), MEXT, Japan, by Japan Society for the Promotion of Science (JSPS) KAKENHI grant No. JP20H05853, and by the UC Southern California Hub with funding from the UC National Laboratories division of the University of California Office of the President. 
Y.I. is supported by JSPS KAKENHI grant Nos. JP18H05458, JP19K14772, and JP22K18277.

\vspace{5mm}

\bibliographystyle{unsrt-author4}
\bibliography{bibtex}

%\clearpage

%\clearpage

%\clearpage

%\clearpage

\clearpage

\begin{table}
\caption{Parameters of AGN LF at 325 MHz \label{tbl:radiolf}}
\begin{tabular}{c @{\qquad}c @{\qquad}c}
\botrule
\textrm{Parameter} & \textrm{PDE} & \textrm{PLE} \\
\colrule
$\log_{10}{(L^*_r/{\rm W \:Hz}^{-1})}$  & 26.26 & 25.96\\
$\log_{10}{(\Phi^*_r/{\rm Mpc}^{-3})}$  & $-$6.40 & $-$6.27\\
$\alpha_r$                              & $-$3.08 & $-$3.02\\
$\beta_r$                               & $-$0.44 & $-$0.44\\
$k_r$                                   & 0.92  & 2.13\\
\botrule
\end{tabular}
\end{table}

\end{document}